\documentclass[twocolumn,showpacs,preprintnumbers,amsmath,amssymb,prl]{revtex4}
\usepackage{graphicx}% Include figure files
\usepackage{dcolumn}% Align table columns on decimal point
\usepackage{bm}% bold math
\usepackage{wasysym}% integrals, etc
\usepackage{subfigure}% subfigures
\usepackage{float}% figuras fixadas

\newcolumntype{C}[1]{>{\centering\let\newline\\\arraybackslash\hspace{0pt}}m{#1}} % column type for table II

\begin{document}

\preprint{}

\title{\textit{Ab initio} quasiparticle bandstructure of ABA and ABC-stacked graphene trilayers}% Force line breaks with \\

\author{Marcos G. Menezes$^1$}
 \email{marcosgm@if.ufrj.br}
\author{Rodrigo B. Capaz$^{1}$}%
\author{Steven G. Louie$^{2,3}$}
\affiliation{$^1$ Instituto de F\'{i}sica, Universidade Federal do Rio de Janeiro, Caixa Postal 68528 21941-972, Rio de Janeiro, RJ, Brazil \\ $^2$ Department of Physics, University of California at Berkeley, Berkeley, CA 94720, USA \\ $^3$ Materials Sciences Division, Laurence Berkeley National Laboratory, Berkeley, CA 94720, USA}

\date{\today}% It is always \today, today,

\begin{abstract}
\b
We obtain the quasiparticle band structure of ABA and ABC-stacked graphene trilayers through \textit{ab initio} density functional theory (DFT) and many-body quasiparticle calculations within the GW approximation. To interpret our results, we fit the DFT and GW $\pi$ bands to a low energy tight-binding model, which is found to reproduce very well the observed features near the K point. The values of the extracted hopping parameters are reported and compared with available theoretical and experimental data. For both stackings, the self energy corrections lead to a renormalization of the Fermi velocity, an effect also observed in previous calculations on monolayer graphene. They also increase the separation between the higher energy bands, which is proportional to the nearest neighbor interlayer hopping parameter $\gamma_1$. Both features are brought to closer agreement with experiment through the self energy corrections. Finally, other effects, such as trigonal warping, electron-hole asymmetry and energy gaps are discussed in terms of the associated parameters.
\end{abstract}

\maketitle

\begin{figure}[h]
\centering
\includegraphics[width=8.5cm]{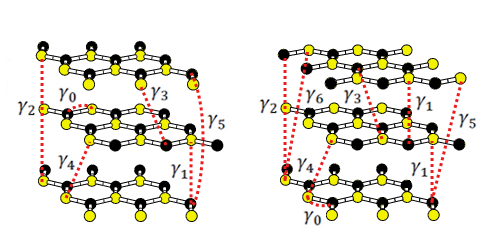}
\caption{\label{geometry} Arrangement of carbon atoms on the ABA (left) and ABC (right) graphene trilayers. Atoms belonging to the A and B sublattices of each layer are represented by yellow and black spheres, respectively. The red dashed lines indicate the tight-binding parameters included in the model used to fit our first-principles results (see text).}
\end{figure}

Graphene, a 2D sheet of carbon atoms in a honeycomb lattice, has attracted a lot of attention of the scientific community in the last few years due to its unique electronic properties, which lead to several potential applications in nanoelectronics \cite{novoselov_science,neto_rmp}. However, since graphene is a zero-gap semiconductor, much of the current effort is directed in finding ways to open a gap for use in electronic devices. In particular, one way to do that is to consider graphene stacks, where a number of layers are stacked on top of each other with a particular arrangement. Much work has been done on bilayer graphene, where it was found that a tunable gap can be opened through application of an external electrical field or through doping \cite{ohta-science,zhang-nature,zhang-prb,neto-prl}. In light of recent experimental progress, graphene trilayers are also attracting increasing attention, revealing electronic properties that depend on the stacking order of the three layers. The two most important stackings are ABA (Bernal) and ABC (rhombohedral), which are shown in Fig. \ref{geometry}. For ABA stacking, the low energy $\pi$ bands are predicted to consist of a set of monolayer and bilayer-like bands, with linear and quadratic dispersions, respectively \cite{neto_rmp,guinea-prb,partoens-prb}. Therefore, this trilayer is expected to show mixed properties from these two systems, which were already observed experimentally \cite{expABA}. In the presence of an external electrical field perpendicular to the layers, these bands hybridize and a tunable overlap between the linear and parabolic bands is introduced \cite{McCannABA,wu-DFT}. In contrast, for ABC stacking, the low energy bands consist of a pair of bands with cubic dispersion, which are very flat near the Fermi level. The large density of states associated with this behavior indicates that many-body interactions might play a crucial role in this case. In fact, there are already a few works in the literature investigating the possibility of different competing phases in this system, such as ferromagnetic order \cite{smith-prb}, charge and spin-density waves and quantum spin Hall phases \cite{scherer-prb}, and even superconductivity \cite{kopnin-prb,peeters-prb-2013}. Moreover, an external electrical field breaks the inversion symmetry of this trilayer and induces a tunable band gap, in a similar fashion to bilayer graphene \cite{guinea-prb,MacDonaldABC,wu-DFT} and also observed experimentally \cite{expABC,HallABC}. We also point out that a similar type of dispersion and an associated quantum critical behavior was observed on a very different system, namely, the Laves phase of $Nb_{1+c}Fe_{2-c}$ \cite{neal-prb,alam-prl}.

In this work, we report first-principle calculations for ABA and ABC-stacked graphene trilayers, employing density functional theory (DFT) \cite{ho-kohn,kohn-sham} and many-body quasiparticle corrections within the GW approximation. Following the framework of Hybertsen and Louie \cite{louie-GW}, our calculations are done in two steps: First, a mean-field step is performed (DFT in our case), where the wavefunctions and energy bands are evaluated and stored. Then, in the next step, the quasiparticle corrections to the DFT energies are calculated within the GW approximation to the electron self energy, by using the DFT energies and wavefunctions \footnote{This one-step procedure is known as a ``$G_oW_o$'' or a ``one-shot'' GW calculation. For more details, see Ref. \cite{louie-GW}}. In our calculations, the DFT step is carried on using the LDA (Perdew-Zunger) approximation for the exchange-correlation functional \cite{perdew-zunger}, Troullier-Martins pseudopotentials for the electron-ion interaction \cite{troullier-martins} and a plane wave basis set to expand the wavefunctions, with an energy cutoff of $60$ Ry, as implemented in the Quantum Espresso package \cite{espresso}. The theoretical lattice constant for this pseudopotential is $a = 2.45$\AA \ and the interlayer distance is set to the experimental value of graphite, $d = 3.35$\AA. We also use a vacuum region of $10.0$\AA \ in the direction perpendicular to the layers in order to avoid interaction between periodic images. In the second step, the many-body calculations are perfomed following the scheme described below, which is implemented in the BerkeleyGW package \cite{BGW}. For the calculation of the dielectric matrix in reciprocal space, we use an energy cutoff of $9.0$ Ry and a coarse $40 \times 40$ Monkhorst-Pack k-point sampling \cite{monkhorst-pack} with $200$ unnocuppied bands for general $\mathbf{q}$ points away from zero and a fine $160 \times 160$ grid and $15$ unnoccupied bands for $\mathbf{q}\rightarrow 0$. This matrix is first evaluated in the static limit by using the RPA approximation and then extended to non-zero frequencies by using a generalized plasmon pole (GPP) model. The self-energy operator $\Sigma$ is evaluated in the same coarse grid. In this step, the Coulomb interaction is truncated in the middle of the vacuum region of the slab. Finally, the bandstructure plots for the LDA and GW bands are generated along high-symmetry lines by using Wannier interpolation, as implemented in the Wannier90 package \cite{wannier90}.

\begin{figure}
\centering
\includegraphics[width=8.5cm]{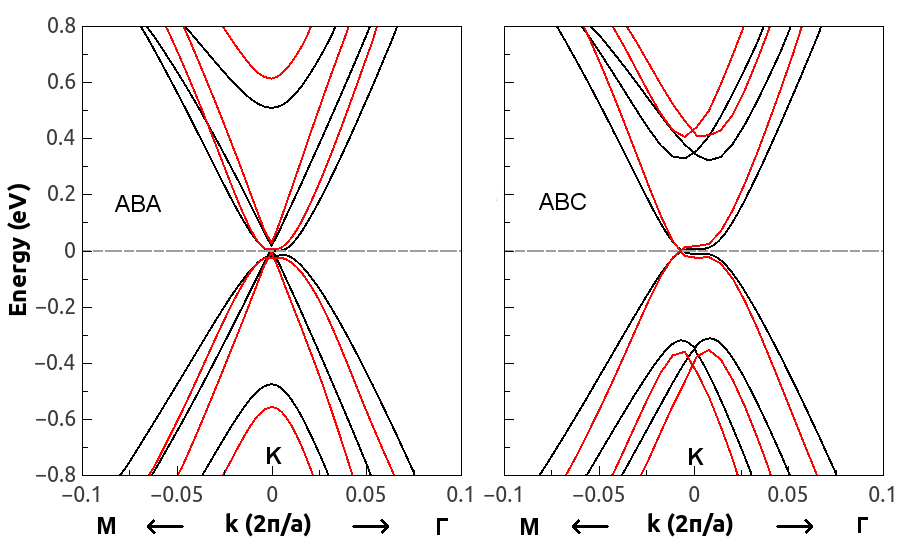}
\caption{\label{bandsab} Comparison between the LDA (black lines) and GW (red lines) bandstructures obtained from our calculations for ABA (left) and ABC (right) stacking. The Brillouin Zone path is along the M-K-$\Gamma$ directions from left to right and it is centered in K. The Fermi level is set to zero in all cases, both in LDA and GW.}
\end{figure}

The results of our calculations are shown in Fig. \ref{bandsab}, where we compare the LDA and GW bandstructures in the energy range of the $\pi$ bands. For both calculations, we see that the bandstructures share the same qualitative features. For ABA stacking (left), the bandstructure near the Fermi level consist of a superposition of a pair of nearly linear bands, resembling those of monolayer graphene, and two pairs of parabolic bands, resembling those of bilayer graphene. However, unlike single and bilayer graphene, both sets of bands have small band gaps, due to the lack of inversion symmetry (or equivalently, A-B sublattice symmetry) in this trilayer. There is also a small offset between the linear and parabolic bands. The values of the gaps at the K point and energy offsets for LDA and GW are reported in Table \ref{gapsABA}, where they are also compared with recent experimental data. The agreement is good, specially for the energy offset. The GW gaps are systematically larger than the corresponding LDA gaps, which is a common trend observed in GW calculations \cite{louie-GW}, and they are also larger than the experimental values. Possible reasons for this discrepancy are the unavoidable residual doping and substrate effects present in experiments, which tend to enhance screening and decrease the quasiparticle gap. We also point out that, although the LDA gaps seem to agree well with the experimental values, such an agreement is only fortuitous. It is well known that DFT underestimates quasiparticle gaps, even if the exchange-correlation functional was known exactly \cite{martin-book}. Another important effect of the quasiparticle corrections is the renormalization of the Fermi velocity, which is visible from the increase of slope of the linear bands in GW when compared to LDA, a feature also observed in previous GW calculations in monolayer graphene \cite{louie-nano,louie-prl,trevisanutto-prl}.

For ABC stacking (right), the bandstructure is very different from the previous case. Near the Fermi level, there is a pair of bands with cubic dispersion, which are very flat near the K point. Moreover, in contrast with ABA, the ABC trilayer does have inversion symmetry, so the valence and conduction bands touch at the Fermi level. Due to the presence of a gap at the K point, these bands touch at three equivalent points located along the $M-K$ lines, where the dispersion is roughly linear. Further away from the Fermi level, there are two pairs of parabolic energy bands. 

\begin{table}
\caption{\label{gapsABA} Energy gaps and energy offset (meV) of the monolayer and bilayer-like bands in the ABA-stacked trilayer. The offset is defined as the energy difference between the middle of the two gaps.}
\begin{tabular}{ l *{3}{p{2cm}} }
\hline \hline
              & LDA & GW & Exp. \cite{expABA} \\
\hline
Monolayer gap & 12  & 35 &  7  \\
Bilayer gap   & 17  & 32 & 14  \\
Offset        & 22  & 21 & 25  \\
\hline \hline
\end{tabular}
\end{table}

Next, we fit the calculated \textit{ab initio} $\pi$ bands to a low energy tight-binding (TB) model in order to extract the parameters that best describe the quasiparticle bandstructures. The hopping parameters included in our model are shown in Fig. \ref{geometry}, where the red dashed lines indicate the atoms connected by them. Atoms belonging to the inequivalent A and B sublattices are represented by yellow and dark spheres, respectively. Following common notation, $\gamma_0$ is the nearest neighbor intralayer hopping and $\gamma_1$ is the nearest neighbor interlayer hopping, connecting atoms that are right on top of each other in adjacent layers. These two parameters are sufficient to describe the main differences observed in the bandstructures, such as the type of dispersion of the low energy bands and their separations. The other parameters, also shown in Fig. \ref{geometry}, describe finer details and follow from a generalization of the Slonczewiski-Weiss-McClure (SWM) model of bulk graphite \cite{dresselhaus} and we use the same definition and sign convention for the parameters that are common to this model, which is extensively discussed in the literature \cite{guinea-prb,partoens-prb,expABA,McCannABA,MacDonaldABC}. Another parameter, $\delta$, not shown in  Fig. 1, corresponds to the onsite energy difference between non-equivalent carbon atoms: High-energy sites correspond to carbon atoms on top of each other (connected by the $\gamma_1$ hopping) and low-energy sites correspond to carbon atoms on top of hexagon centers in adjacent layers. Finally, we point out that surface effects are neglected in our TB calculations, since they were found to be very small in previous calculations on the ABA trilayer \cite{McCannABA}.

\begin{figure}
\centering
\includegraphics[width=8.5cm]{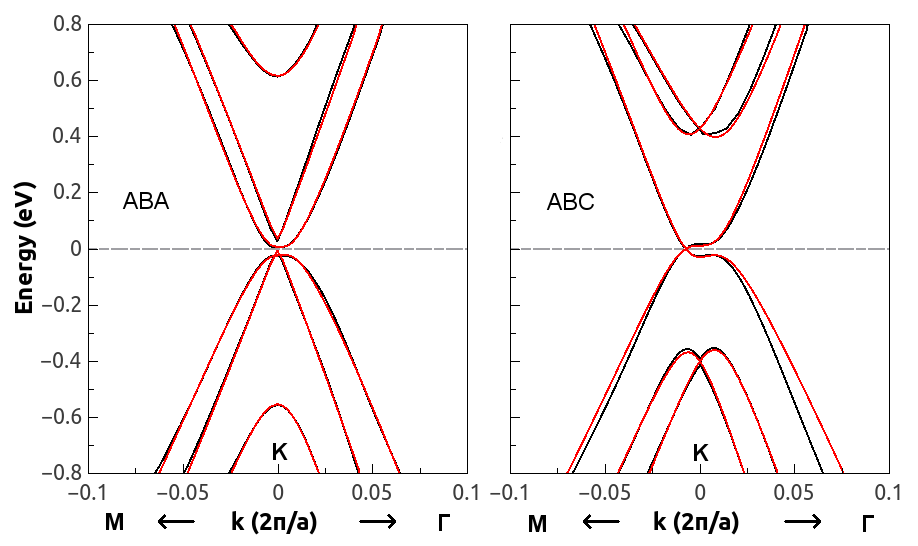}
\caption{\label{bandsfit} Comparison between the quasiparticle bands (black lines) and the corresponding adjusted TB bands (red lines) for ABA (left) and ABC (right) stacking. The path is the same as in Fig. \ref{bandsab} and the Fermi level is set to zero in both cases. Fitted parameters are shown in Table \ref{param}.}
\end{figure}

We employ the least-squares method to fit the TB bands to either LDA or GW $\pi$ bands from our calculations, thus obtaining the set of parameters that best describes them. Since we want to describe the details of the bands near the K point (such as the small energy gaps), only k-points within a radius of $0.02 \ 2\pi/a$ from the K point are included in our fits. The comparison between GW bands and TB bands is shown in Fig. \ref{bandsfit}. We can see that the TB model describes very well the features observed in our calculations, even outside the radius of the fit. The fitted parameters are reported in Table \ref{param}. Different sign conventions are used in the literature, so we explicitly indicate when a different convention is being used. For both stackings, the quasiparticle corrections increase the value of $\gamma_0$, renormalizing the Fermi velocity ($v_0 = \sqrt{3}\gamma_0 a/2$) by about $28\%$ in ABA and $24\%$ in ABC trilayer, with respect to the LDA values. The GW value is in agreement with previous GW calculations on monolayer graphene \cite{louie-nano,louie-prl,trevisanutto-prl} and with the experimental value for the monolayer \cite{zhang-fermivel}. The parameter $\gamma_1$, which is associated with the distance between the higher energy bands and the Fermi level, is also increased by the quasiparticle corrections. In the absence of electron-hole asymmetry, these bands have energies $\pm \sqrt{2}\gamma_1$ for ABA and $\pm \gamma_1$ for ABC stacking at the K point \cite{McCannABA}.

\begin{table*}
\caption{\label{param} Tight-binding parameters (in eV) obtained from our calculations (LDA or GW columns) and comparison with values from the literature, both theoretical and experimental. Wherever no value is shown, the corresponding parameter is either not applicable to that case or is set to zero. $\sigma$ (in eV) is the standard deviation (error bars) of the adjustment inside its range.}
\begin{tabular*}{\textwidth}
%{ *{4}{p{1.85cm}} | *{3}{p{1.85cm}} | *{2}{p{1.85cm}}}
%16.5cm
{p{1.85cm} *{3}{C{1.85cm}} | *{3}{C{1.85cm}} | *{2}{C{1.85cm}} }
\hline \hline
           &     & ABA &      &     & ABC &     & Bilayer & Graphite \\
Parameter  & LDA & GW & Exp. \cite{expABA} & LDA & GW & Theor. \cite{MacDonaldABC} & Exp. \cite{zhang-prb} & Exp. \cite{dresselhaus} \\
\hline
$\gamma_0$ &  2.590 &  3.306  &  3.100$^a$ &  2.577 &  3.188  &  3.160$^a$ & 3.000$^a$ &  3.160 \\
$\gamma_1$ &  0.348 &  0.414  &  0.390$^a$ &  0.348 &  0.415  &  0.502     & 0.400     &  0.390 \\
$\gamma_2$ & -0.043 & -0.060  & -0.028     & -0.024 & -0.041  & -0.017     &   -       & -0.020 \\
$\gamma_3$ &  0.283 &  0.242  &  0.315$^a$ &  0.290 &  0.323  &  0.377$^b$ & 0.300$^a$ &  0.315 \\
$\gamma_4$ &  0.162 &  0.152  &  0.041     &  0.196 &  0.287  &  0.099$^b$ & 0.150     &  0.044 \\
$\gamma_5$ &  0.024 &  0.052  &  0.050     &  0.019 &  0.126  &    -       &   -       &  0.038 \\
$\gamma_6$ &    -   &    -    &    -       &  0.004 &  0.084  &    -       &   -       &   -    \\
$\delta$   &  0.010 &  0.012  &  0.046     &  0.001 &  0.023  &  0.001$^b$ & 0.018     &  0.050$^c$ \\
$\sigma$   &  0.002 &  0.006  &    -       &  0.002 &  0.006  &    -       &   -       &   -    \\
\hline \hline
\multicolumn{9}{l}{$^a$ This parameter was not obtained in this reference, it was set to a value from the literature.} \\
\multicolumn{9}{l}{$^b$ The sign of this parameter was changed from the original reference in order to match our convention.} \\
\multicolumn{9}{l}{$^c$ $\delta$ should not be confused with $\Delta = \delta + \gamma_2 - \gamma_5 = -0.008$ eV, a similar parameter used in graphite.}
\end{tabular*}
\end{table*}

We now discuss in detail the effects of the remaining TB parameters to the bandstructure features. The parameter $\gamma_3$ causes a trigonal distortion of the low energy bands. As can be seen in Fig. \ref{bandsfit}, in ABA stacking, the parabolic bands have four energy minima: one at the K point and three along the equivalent $\Gamma-K$ lines. In our model, this is consistent with a positive sign for $\gamma_3$ \footnote{ Note that it's common to find an opposite sign convention for $\gamma_3$ in the literature \cite{MacDonaldABC,McCannABA,McCann-review,expABA}, so care must be taken when comparing results and using reference values. A discussion of the conventions and their relation to the SWM model can be found in Ref. \cite{partoens-prb}.}, in agreement with bulk graphite and bilayer graphene. In ABC, $\gamma_3$ has a similar effect and is also positive, but there is an additional distortion caused by $\gamma_2$, as we discuss below. The parameter $\gamma_4$ is responsible for a small electron-hole asymmetry in the bands. The ABA values for this parameter agree better with the experimental value from Bernal bilayer graphene than with the experimental value for this trilayer \cite{expABA,zhang-prb} (Table \ref{param}). On the other hand, the ABC values appear to be overestimated, since the TB bands show a larger asymmetry than predicted by LDA and GW, specially outside the range of the fit. This could be related to an insensitivity of the fit to this parameter, that is, since $\gamma_4$ changes the curvature of the bands and they are mostly flat in the range of the adjustment, the curvature change may be not being correctly reproduced outside that range.

The remaining parameters have quite different roles for each trilayer. For ABA stacking, the parameters $\gamma_2$, $\gamma_5$ and $\delta$ describe the inequivalency between A and B sublattices and are responsible for the opening of small band gaps in the linear and parabolic bands, also introducing an offset between them. As can be seen in Fig. \ref{bandsfit}, the TB model reproduces very well these features and the adjusted parameters are in good agreement with values from the literature. For ABC stacking, the parameter $\gamma_2$ also opens a gap between the cubic bands at the K point, but the presence of inversion symmetry prevents the full opening of a band gap in this case. Hence, $\gamma_2$ also induces a trigonal distortion, shifting the touching points of these bands from the K point to three points at the equivalent $M-K$ lines. This is consistent with a negative sign for $\gamma_2$ in our model, in agreement with ABA trilayer and graphite. On the other hand, the parameters $\gamma_5$ and $\gamma_6$ don't have any visible effects on the ABC bandstructure in the range considered, so they can be safely discarded in the description of the low energy bands, as was done previously \cite{MacDonaldABC}. Nevertheless, we include in Table \ref{param} the values given by the adjustment for future reference, but we stress that they could be also being affected by insensitivity. Note that the ABC value for $\gamma_5$ cannot be directly compared with the values for ABA and graphite, since they have different definitions and roles. The parameter $\delta$ also plays a small role in ABC trilayer, introducing only a small electron-hole asymmetry in a similar fashion to $\gamma_4$. This could explain why the LDA value is smaller than the corresponding value for ABA, in agreement with the value obtained in a previous DFT calculation \cite{MacDonaldABC}. On the other hand, the GW value is larger than the corresponding ABA value, which could indicate either a larger asymmetry of the bands or again an insensitivity of the adjustment.

In conclusion, we have studied the quasiparticle band structure of ABA and ABC trilayer graphene within the GW approximation. The $\pi$ bands obtained from these calculations were fitted to a TB model and the corresponding parameters were extracted. This model is found to reproduce very well the observed features in all cases, such as the type of dispersions and energy gaps. The fit parameters show a good agreement with available data from graphene bilayers, trilayers and graphite, and finer details of the bands are also correctly reproduced by them in the range of the fitting. The main effects of the quasiparticle corrections are the renormalization of the Fermi velocity and the increase of the separation between the higher energy bands, where both bring theory to closer agreement with experiment. Therefore, we expect this work to provide future reference for studies on graphene trilayers and other stackings, which can reveal more interesting properties and applications.

\begin{acknowledgments}
We thank Felipe Jornada for valuable discussions. This work was supported by the Brazilian funding agencies: CAPES, CNPq, FAPERJ and INCT-Nanomateriais de Carbono. Steven G. Louie acknowledges support from a Simons Foundation Fellowship in Theoretical Physics, and support from National Science Foundation Grant No. DMR10-1006184 (DFT and tight-binding analysis) and the Director, Office of Science, Office Basic Energy Sciences, Materials Sciences and Engineering Division, U.S. Department of Energy under contract No. DE-AC02-05CH11231 (GW simulations). We also thank NERSC for the computational resources employed in our calculations.
\end{acknowledgments}

\bibliography{Artigo}% Produces the bibliography via BibTeX.

\begin{thebibliography}{36}
\expandafter\ifx\csname natexlab\endcsname\relax\def\natexlab#1{#1}\fi
\expandafter\ifx\csname bibnamefont\endcsname\relax
  \def\bibnamefont#1{#1}\fi
\expandafter\ifx\csname bibfnamefont\endcsname\relax
  \def\bibfnamefont#1{#1}\fi
\expandafter\ifx\csname citenamefont\endcsname\relax
  \def\citenamefont#1{#1}\fi
\expandafter\ifx\csname url\endcsname\relax
  \def\url#1{\texttt{#1}}\fi
\expandafter\ifx\csname urlprefix\endcsname\relax\def\urlprefix{URL }\fi
\providecommand{\bibinfo}[2]{#2}
\providecommand{\eprint}[2][]{\url{#2}}

\bibitem[{\citenamefont{Novoselov et~al.}(2004)\citenamefont{Novoselov, Geim,
  Morozov, Jiang, Zhang, Dubonos, Grigorieva, and Firsov}}]{novoselov_science}
\bibinfo{author}{\bibfnamefont{K.~S.} \bibnamefont{Novoselov}},
  \bibinfo{author}{\bibfnamefont{A.~K.} \bibnamefont{Geim}},
  \bibinfo{author}{\bibfnamefont{S.~V.} \bibnamefont{Morozov}},
  \bibinfo{author}{\bibfnamefont{D.}~\bibnamefont{Jiang}},
  \bibinfo{author}{\bibfnamefont{Y.}~\bibnamefont{Zhang}},
  \bibinfo{author}{\bibfnamefont{S.~V.} \bibnamefont{Dubonos}},
  \bibinfo{author}{\bibfnamefont{I.~V.} \bibnamefont{Grigorieva}},
  \bibnamefont{and} \bibinfo{author}{\bibfnamefont{A.~A.}
  \bibnamefont{Firsov}}, \bibinfo{journal}{Science}
  \textbf{\bibinfo{volume}{305}}, \bibinfo{pages}{666} (\bibinfo{year}{2004}).

\bibitem[{\citenamefont{Neto et~al.}(2009)\citenamefont{Neto, Guinea, Peres,
  Novoselov, and Geim}}]{neto_rmp}
\bibinfo{author}{\bibfnamefont{A.~H.~C.} \bibnamefont{Neto}},
  \bibinfo{author}{\bibfnamefont{F.}~\bibnamefont{Guinea}},
  \bibinfo{author}{\bibfnamefont{N.~M.~R.} \bibnamefont{Peres}},
  \bibinfo{author}{\bibfnamefont{K.~S.} \bibnamefont{Novoselov}},
  \bibnamefont{and} \bibinfo{author}{\bibfnamefont{A.~K.} \bibnamefont{Geim}},
  \bibinfo{journal}{Rev. Mod. Phys.} \textbf{\bibinfo{volume}{81}},
  \bibinfo{pages}{109} (\bibinfo{year}{2009}).

\bibitem[{\citenamefont{Ohta et~al.}(2006)\citenamefont{Ohta, Bostwick,
  Seyller, Horn, and Rotenberg}}]{ohta-science}
\bibinfo{author}{\bibfnamefont{T.}~\bibnamefont{Ohta}},
  \bibinfo{author}{\bibfnamefont{A.}~\bibnamefont{Bostwick}},
  \bibinfo{author}{\bibfnamefont{T.}~\bibnamefont{Seyller}},
  \bibinfo{author}{\bibfnamefont{K.}~\bibnamefont{Horn}}, \bibnamefont{and}
  \bibinfo{author}{\bibfnamefont{E.}~\bibnamefont{Rotenberg}},
  \bibinfo{journal}{Science} \textbf{\bibinfo{volume}{313}},
  \bibinfo{pages}{951} (\bibinfo{year}{2006}).

\bibitem[{\citenamefont{Zhang et~al.}(2009)\citenamefont{Zhang, Tang, Girit,
  Hao, Martin, Zettl, Crommie, Shen, and Wang}}]{zhang-nature}
\bibinfo{author}{\bibfnamefont{Y.}~\bibnamefont{Zhang}},
  \bibinfo{author}{\bibfnamefont{T.-T.} \bibnamefont{Tang}},
  \bibinfo{author}{\bibfnamefont{C.}~\bibnamefont{Girit}},
  \bibinfo{author}{\bibfnamefont{Z.}~\bibnamefont{Hao}},
  \bibinfo{author}{\bibfnamefont{M.~C.} \bibnamefont{Martin}},
  \bibinfo{author}{\bibfnamefont{A.}~\bibnamefont{Zettl}},
  \bibinfo{author}{\bibfnamefont{M.~F.} \bibnamefont{Crommie}},
  \bibinfo{author}{\bibfnamefont{Y.~R.} \bibnamefont{Shen}}, \bibnamefont{and}
  \bibinfo{author}{\bibfnamefont{F.}~\bibnamefont{Wang}},
  \bibinfo{journal}{Nature Lett.} \textbf{\bibinfo{volume}{459}},
  \bibinfo{pages}{820} (\bibinfo{year}{2009}).

\bibitem[{\citenamefont{Zhang et~al.}(2008)\citenamefont{Zhang, Li, Basov,
  Fogler, Hao, and Martin}}]{zhang-prb}
\bibinfo{author}{\bibfnamefont{L.~M.} \bibnamefont{Zhang}},
  \bibinfo{author}{\bibfnamefont{Z.~Q.} \bibnamefont{Li}},
  \bibinfo{author}{\bibfnamefont{D.~N.} \bibnamefont{Basov}},
  \bibinfo{author}{\bibfnamefont{M.~M.} \bibnamefont{Fogler}},
  \bibinfo{author}{\bibfnamefont{Z.}~\bibnamefont{Hao}}, \bibnamefont{and}
  \bibinfo{author}{\bibfnamefont{M.~C.} \bibnamefont{Martin}},
  \bibinfo{journal}{Phys. Rev. B} \textbf{\bibinfo{volume}{78}},
  \bibinfo{pages}{235408} (\bibinfo{year}{2008}).

\bibitem[{\citenamefont{Castro et~al.}(2007)\citenamefont{Castro, Novoselov,
  Morozov, Peres, dos Santos, Nilsson, Guinea, Geim, and Neto}}]{neto-prl}
\bibinfo{author}{\bibfnamefont{E.~V.} \bibnamefont{Castro}},
  \bibinfo{author}{\bibfnamefont{K.~S.} \bibnamefont{Novoselov}},
  \bibinfo{author}{\bibfnamefont{S.}~\bibnamefont{Morozov}},
  \bibinfo{author}{\bibfnamefont{N.~M.~R.} \bibnamefont{Peres}},
  \bibinfo{author}{\bibfnamefont{J.~M. B.~L.} \bibnamefont{dos Santos}},
  \bibinfo{author}{\bibfnamefont{J.}~\bibnamefont{Nilsson}},
  \bibinfo{author}{\bibfnamefont{F.}~\bibnamefont{Guinea}},
  \bibinfo{author}{\bibfnamefont{A.~K.} \bibnamefont{Geim}}, \bibnamefont{and}
  \bibinfo{author}{\bibfnamefont{A.~H.~C.} \bibnamefont{Neto}},
  \bibinfo{journal}{Phys. Rev. Lett.} \textbf{\bibinfo{volume}{99}},
  \bibinfo{pages}{216802} (\bibinfo{year}{2007}).

\bibitem[{\citenamefont{Guinea et~al.}(2006)\citenamefont{Guinea, Neto, and
  Peres}}]{guinea-prb}
\bibinfo{author}{\bibfnamefont{F.}~\bibnamefont{Guinea}},
  \bibinfo{author}{\bibfnamefont{A.~H.~C.} \bibnamefont{Neto}},
  \bibnamefont{and} \bibinfo{author}{\bibfnamefont{N.~M.~R.}
  \bibnamefont{Peres}}, \bibinfo{journal}{Phys. Rev. B}
  \textbf{\bibinfo{volume}{73}}, \bibinfo{pages}{245426}
  (\bibinfo{year}{2006}).

\bibitem[{\citenamefont{Partoens and Peeters}(2006)}]{partoens-prb}
\bibinfo{author}{\bibfnamefont{B.}~\bibnamefont{Partoens}} \bibnamefont{and}
  \bibinfo{author}{\bibfnamefont{F.~M.} \bibnamefont{Peeters}},
  \bibinfo{journal}{Phys. Rev. B} \textbf{\bibinfo{volume}{74}},
  \bibinfo{pages}{075404} (\bibinfo{year}{2006}).

\bibitem[{\citenamefont{Taychatanapat et~al.}(2011)\citenamefont{Taychatanapat,
  Watanabe, Taniguchi, and Jarillo-Herrero}}]{expABA}
\bibinfo{author}{\bibfnamefont{T.}~\bibnamefont{Taychatanapat}},
  \bibinfo{author}{\bibfnamefont{K.}~\bibnamefont{Watanabe}},
  \bibinfo{author}{\bibfnamefont{T.}~\bibnamefont{Taniguchi}},
  \bibnamefont{and}
  \bibinfo{author}{\bibfnamefont{P.}~\bibnamefont{Jarillo-Herrero}},
  \bibinfo{journal}{Nature Phys.} \textbf{\bibinfo{volume}{7}},
  \bibinfo{pages}{621} (\bibinfo{year}{2011}).

\bibitem[{\citenamefont{Koshino and McCann}(2009)}]{McCannABA}
\bibinfo{author}{\bibfnamefont{M.}~\bibnamefont{Koshino}} \bibnamefont{and}
  \bibinfo{author}{\bibfnamefont{E.}~\bibnamefont{McCann}},
  \bibinfo{journal}{Phys. Rev. B} \textbf{\bibinfo{volume}{79}},
  \bibinfo{pages}{125443} (\bibinfo{year}{2009}).

\bibitem[{\citenamefont{Wu}(2011)}]{wu-DFT}
\bibinfo{author}{\bibfnamefont{B.-R.} \bibnamefont{Wu}},
  \bibinfo{journal}{Appl. Phys. Lett.} \textbf{\bibinfo{volume}{98}},
  \bibinfo{pages}{263107} (\bibinfo{year}{2011}).

\bibitem[{\citenamefont{Olsen et~al.}(2013)\citenamefont{Olsen, van Gelderen,
  and Smith}}]{smith-prb}
\bibinfo{author}{\bibfnamefont{R.}~\bibnamefont{Olsen}},
  \bibinfo{author}{\bibfnamefont{R.}~\bibnamefont{van Gelderen}},
  \bibnamefont{and} \bibinfo{author}{\bibfnamefont{C.~M.} \bibnamefont{Smith}},
  \bibinfo{journal}{Phys. Rev. B} \textbf{\bibinfo{volume}{87}},
  \bibinfo{pages}{115414} (\bibinfo{year}{2013}).

\bibitem[{\citenamefont{Scherer et~al.}(2012)\citenamefont{Scherer, Uebelacker,
  Scherer, and Honerkamp}}]{scherer-prb}
\bibinfo{author}{\bibfnamefont{M.~M.} \bibnamefont{Scherer}},
  \bibinfo{author}{\bibfnamefont{S.}~\bibnamefont{Uebelacker}},
  \bibinfo{author}{\bibfnamefont{D.~D.} \bibnamefont{Scherer}},
  \bibnamefont{and}
  \bibinfo{author}{\bibfnamefont{C.}~\bibnamefont{Honerkamp}},
  \bibinfo{journal}{Phys. Rev. B} \textbf{\bibinfo{volume}{86}},
  \bibinfo{pages}{155415} (\bibinfo{year}{2012}).

\bibitem[{\citenamefont{Kopnin et~al.}(2011)\citenamefont{Kopnin,
  T.~T.~Heikkil{\"a}, and Volovik}}]{kopnin-prb}
\bibinfo{author}{\bibfnamefont{N.~B.} \bibnamefont{Kopnin}},
  \bibinfo{author}{\bibfnamefont{.}~\bibnamefont{T.~T.~Heikkil{\"a}}},
  \bibnamefont{and} \bibinfo{author}{\bibfnamefont{G.~E.}
  \bibnamefont{Volovik}}, \bibinfo{journal}{Phys. Rev. B}
  \textbf{\bibinfo{volume}{83}}, \bibinfo{pages}{220503(R)}
  (\bibinfo{year}{2011}).

\bibitem[{\citenamefont{Mu{\~n}oz et~al.}(2013)\citenamefont{Mu{\~n}oz, Covaci,
  and Peeters}}]{peeters-prb-2013}
\bibinfo{author}{\bibfnamefont{W.~A.} \bibnamefont{Mu{\~n}oz}},
  \bibinfo{author}{\bibfnamefont{L.}~\bibnamefont{Covaci}}, \bibnamefont{and}
  \bibinfo{author}{\bibfnamefont{F.~M.} \bibnamefont{Peeters}},
  \bibinfo{journal}{Phys. Rev. B} \textbf{\bibinfo{volume}{87}},
  \bibinfo{pages}{134509} (\bibinfo{year}{2013}).

\bibitem[{\citenamefont{Zhang et~al.}(2010)\citenamefont{Zhang, Sahu, Min, and
  MacDonald}}]{MacDonaldABC}
\bibinfo{author}{\bibfnamefont{F.}~\bibnamefont{Zhang}},
  \bibinfo{author}{\bibfnamefont{B.}~\bibnamefont{Sahu}},
  \bibinfo{author}{\bibfnamefont{H.}~\bibnamefont{Min}}, \bibnamefont{and}
  \bibinfo{author}{\bibfnamefont{A.~H.} \bibnamefont{MacDonald}},
  \bibinfo{journal}{Phys. Rev. B} \textbf{\bibinfo{volume}{82}},
  \bibinfo{pages}{035409} (\bibinfo{year}{2010}).

\bibitem[{\citenamefont{Lui et~al.}(2011)\citenamefont{Lui, Li, Mak,
  Cappelluti, and Heinz}}]{expABC}
\bibinfo{author}{\bibfnamefont{C.~H.} \bibnamefont{Lui}},
  \bibinfo{author}{\bibfnamefont{Z.}~\bibnamefont{Li}},
  \bibinfo{author}{\bibfnamefont{K.~F.} \bibnamefont{Mak}},
  \bibinfo{author}{\bibfnamefont{E.}~\bibnamefont{Cappelluti}},
  \bibnamefont{and} \bibinfo{author}{\bibfnamefont{T.~F.} \bibnamefont{Heinz}},
  \bibinfo{journal}{Nature Phys.} \textbf{\bibinfo{volume}{7}},
  \bibinfo{pages}{944} (\bibinfo{year}{2011}).

\bibitem[{\citenamefont{Zhang et~al.}(2011)\citenamefont{Zhang, Zhang, Camacho,
  Khodas, and Zaliznyak}}]{HallABC}
\bibinfo{author}{\bibfnamefont{L.}~\bibnamefont{Zhang}},
  \bibinfo{author}{\bibfnamefont{Y.}~\bibnamefont{Zhang}},
  \bibinfo{author}{\bibfnamefont{J.}~\bibnamefont{Camacho}},
  \bibinfo{author}{\bibfnamefont{M.}~\bibnamefont{Khodas}}, \bibnamefont{and}
  \bibinfo{author}{\bibfnamefont{I.~A.} \bibnamefont{Zaliznyak}},
  \bibinfo{journal}{Nature Phys.} \textbf{\bibinfo{volume}{7}},
  \bibinfo{pages}{953} (\bibinfo{year}{2011}).

\bibitem[{\citenamefont{Neal et~al.}(2011)\citenamefont{Neal, Ylvisaker, and
  Pickett}}]{neal-prb}
\bibinfo{author}{\bibfnamefont{B.~P.} \bibnamefont{Neal}},
  \bibinfo{author}{\bibfnamefont{E.~R.} \bibnamefont{Ylvisaker}},
  \bibnamefont{and} \bibinfo{author}{\bibfnamefont{W.~E.}
  \bibnamefont{Pickett}}, \bibinfo{journal}{Phys. Rev. B}
  \textbf{\bibinfo{volume}{84}}, \bibinfo{pages}{085133}
  (\bibinfo{year}{2011}).

\bibitem[{\citenamefont{Alam and Johnson}(2011)}]{alam-prl}
\bibinfo{author}{\bibfnamefont{A.}~\bibnamefont{Alam}} \bibnamefont{and}
  \bibinfo{author}{\bibfnamefont{D.~D.} \bibnamefont{Johnson}},
  \bibinfo{journal}{Phys. Rev. Lett.} \textbf{\bibinfo{volume}{107}},
  \bibinfo{pages}{206401} (\bibinfo{year}{2011}).

\bibitem[{\citenamefont{Hohenberg and Kohn}(1964)}]{ho-kohn}
\bibinfo{author}{\bibfnamefont{P.}~\bibnamefont{Hohenberg}} \bibnamefont{and}
  \bibinfo{author}{\bibfnamefont{W.}~\bibnamefont{Kohn}},
  \bibinfo{journal}{Phys. Rev.} \textbf{\bibinfo{volume}{136}},
  \bibinfo{pages}{B864} (\bibinfo{year}{1964}).

\bibitem[{\citenamefont{Kohn and Sham}(1965)}]{kohn-sham}
\bibinfo{author}{\bibfnamefont{W.}~\bibnamefont{Kohn}} \bibnamefont{and}
  \bibinfo{author}{\bibfnamefont{L.~J.} \bibnamefont{Sham}},
  \bibinfo{journal}{Phys. Rev.} \textbf{\bibinfo{volume}{140}},
  \bibinfo{pages}{A1133} (\bibinfo{year}{1965}).

\bibitem[{\citenamefont{Hybertsen and Louie}(1986)}]{louie-GW}
\bibinfo{author}{\bibfnamefont{M.~S.} \bibnamefont{Hybertsen}}
  \bibnamefont{and} \bibinfo{author}{\bibfnamefont{S.~G.} \bibnamefont{Louie}},
  \bibinfo{journal}{Phys. Rev. B} \textbf{\bibinfo{volume}{34}},
  \bibinfo{pages}{5390} (\bibinfo{year}{1986}).

\bibitem[{\citenamefont{Perdew and Zunger}(1981)}]{perdew-zunger}
\bibinfo{author}{\bibfnamefont{J.~P.} \bibnamefont{Perdew}} \bibnamefont{and}
  \bibinfo{author}{\bibfnamefont{A.}~\bibnamefont{Zunger}},
  \bibinfo{journal}{Phys. Rev. B} \textbf{\bibinfo{volume}{23}},
  \bibinfo{pages}{5048} (\bibinfo{year}{1981}).

\bibitem[{\citenamefont{Troullier and Martins}(1991)}]{troullier-martins}
\bibinfo{author}{\bibfnamefont{N.}~\bibnamefont{Troullier}} \bibnamefont{and}
  \bibinfo{author}{\bibfnamefont{J.~L.} \bibnamefont{Martins}},
  \bibinfo{journal}{Phys. Rev. B} \textbf{\bibinfo{volume}{43}},
  \bibinfo{pages}{1993} (\bibinfo{year}{1991}).

\bibitem[{\citenamefont{et~al}(2009)}]{espresso}
\bibinfo{author}{\bibfnamefont{P.~G.} \bibnamefont{et~al}},
  \bibinfo{journal}{J.Phys.:Condens.Matter} \textbf{\bibinfo{volume}{21}},
  \bibinfo{pages}{395502} (\bibinfo{year}{2009}).

\bibitem[{\citenamefont{Deslippe et~al.}(2012)\citenamefont{Deslippe,
  Samsonidze, Strubbe, Jain, Cohen, and Louie}}]{BGW}
\bibinfo{author}{\bibfnamefont{J.}~\bibnamefont{Deslippe}},
  \bibinfo{author}{\bibfnamefont{G.}~\bibnamefont{Samsonidze}},
  \bibinfo{author}{\bibfnamefont{D.~A.} \bibnamefont{Strubbe}},
  \bibinfo{author}{\bibfnamefont{M.}~\bibnamefont{Jain}},
  \bibinfo{author}{\bibfnamefont{M.~L.} \bibnamefont{Cohen}}, \bibnamefont{and}
  \bibinfo{author}{\bibfnamefont{S.~G.} \bibnamefont{Louie}},
  \bibinfo{journal}{Comput. Phys. Commun.} \textbf{\bibinfo{volume}{183}},
  \bibinfo{pages}{1269} (\bibinfo{year}{2012}).

\bibitem[{\citenamefont{Monkhorst and Pack}(1976)}]{monkhorst-pack}
\bibinfo{author}{\bibfnamefont{H.~J.} \bibnamefont{Monkhorst}}
  \bibnamefont{and} \bibinfo{author}{\bibfnamefont{J.~D.} \bibnamefont{Pack}},
  \bibinfo{journal}{Phys. Rev. B} \textbf{\bibinfo{volume}{13}},
  \bibinfo{pages}{5188} (\bibinfo{year}{1976}).

\bibitem[{\citenamefont{Mostofi et~al.}(2008)\citenamefont{Mostofi, Yates, Lee,
  Souza, Vanderbilt, and Marzari}}]{wannier90}
\bibinfo{author}{\bibfnamefont{A.~A.} \bibnamefont{Mostofi}},
  \bibinfo{author}{\bibfnamefont{J.~R.} \bibnamefont{Yates}},
  \bibinfo{author}{\bibfnamefont{Y.-S.} \bibnamefont{Lee}},
  \bibinfo{author}{\bibfnamefont{I.}~\bibnamefont{Souza}},
  \bibinfo{author}{\bibfnamefont{D.}~\bibnamefont{Vanderbilt}},
  \bibnamefont{and} \bibinfo{author}{\bibfnamefont{N.}~\bibnamefont{Marzari}},
  \bibinfo{journal}{Comput. Phys. Commun.} \textbf{\bibinfo{volume}{178}},
  \bibinfo{pages}{685} (\bibinfo{year}{2008}).

\bibitem[{\citenamefont{Martin}(2004)}]{martin-book}
\bibinfo{author}{\bibfnamefont{R.~M.} \bibnamefont{Martin}},
  \emph{\bibinfo{title}{Electronic Structure: Basic Theory and Practical
  Methods}} (\bibinfo{publisher}{Cambridge University Press},
  \bibinfo{year}{2004}).

\bibitem[{\citenamefont{Park et~al.}(2009)\citenamefont{Park, Giustino,
  Spataru, Cohen, and Louie}}]{louie-nano}
\bibinfo{author}{\bibfnamefont{C.-H.} \bibnamefont{Park}},
  \bibinfo{author}{\bibfnamefont{F.}~\bibnamefont{Giustino}},
  \bibinfo{author}{\bibfnamefont{C.~D.} \bibnamefont{Spataru}},
  \bibinfo{author}{\bibfnamefont{M.~L.} \bibnamefont{Cohen}}, \bibnamefont{and}
  \bibinfo{author}{\bibfnamefont{S.~G.} \bibnamefont{Louie}},
  \bibinfo{journal}{Nano Lett.} \textbf{\bibinfo{volume}{9}},
  \bibinfo{pages}{4234} (\bibinfo{year}{2009}).

\bibitem[{\citenamefont{Yang et~al.}(2009)\citenamefont{Yang, Deslippe, Park,
  Cohen, and Louie}}]{louie-prl}
\bibinfo{author}{\bibfnamefont{L.}~\bibnamefont{Yang}},
  \bibinfo{author}{\bibfnamefont{J.}~\bibnamefont{Deslippe}},
  \bibinfo{author}{\bibfnamefont{C.-H.} \bibnamefont{Park}},
  \bibinfo{author}{\bibfnamefont{M.~L.} \bibnamefont{Cohen}}, \bibnamefont{and}
  \bibinfo{author}{\bibfnamefont{S.~G.} \bibnamefont{Louie}},
  \bibinfo{journal}{Phys. Rev. Lett.} \textbf{\bibinfo{volume}{103}},
  \bibinfo{pages}{186802} (\bibinfo{year}{2009}).

\bibitem[{\citenamefont{Trevisanutto et~al.}(2008)\citenamefont{Trevisanutto,
  Giorgetti, Reining, Ladisa, and Olevano}}]{trevisanutto-prl}
\bibinfo{author}{\bibfnamefont{P.~E.} \bibnamefont{Trevisanutto}},
  \bibinfo{author}{\bibfnamefont{C.}~\bibnamefont{Giorgetti}},
  \bibinfo{author}{\bibfnamefont{L.}~\bibnamefont{Reining}},
  \bibinfo{author}{\bibfnamefont{M.}~\bibnamefont{Ladisa}}, \bibnamefont{and}
  \bibinfo{author}{\bibfnamefont{V.}~\bibnamefont{Olevano}},
  \bibinfo{journal}{Phys. Rev. Lett.} \textbf{\bibinfo{volume}{101}},
  \bibinfo{pages}{226405} (\bibinfo{year}{2008}).

\bibitem[{\citenamefont{Dresselhaus and Dresselhaus}(1981)}]{dresselhaus}
\bibinfo{author}{\bibfnamefont{M.~S.} \bibnamefont{Dresselhaus}}
  \bibnamefont{and}
  \bibinfo{author}{\bibfnamefont{G.}~\bibnamefont{Dresselhaus}},
  \bibinfo{journal}{Advances in Physics} \textbf{\bibinfo{volume}{30}},
  \bibinfo{pages}{139} (\bibinfo{year}{1981}).

\bibitem[{\citenamefont{et~al.}(2005)}]{zhang-fermivel}
\bibinfo{author}{\bibfnamefont{Y.~Z.} \bibnamefont{et~al.}},
  \bibinfo{journal}{Nature (London)} \textbf{\bibinfo{volume}{438}},
  \bibinfo{pages}{201} (\bibinfo{year}{2005}).

\bibitem[{\citenamefont{McCann and Koshino}(2012)}]{McCann-review}
\bibinfo{author}{\bibfnamefont{E.}~\bibnamefont{McCann}} \bibnamefont{and}
  \bibinfo{author}{\bibfnamefont{M.}~\bibnamefont{Koshino}},
  \bibinfo{journal}{cond-mat.mes-hall}  (\bibinfo{year}{2012}).

\end{thebibliography}

\end{document}